\documentclass[conference]{IEEEtran}
\IEEEoverridecommandlockouts

\usepackage{cite}
\usepackage{amsmath,amssymb,amsfonts}
\usepackage{algorithmic}
\usepackage{graphicx}
\usepackage{textcomp}
\usepackage{balance}
\usepackage{xcolor}
\def\BibTeX{{\rm B\kern-.05em{\sc i\kern-.025em b}\kern-.08em
    T\kern-.1667em\lower.7ex\hbox{E}\kern-.125emX}}
\begin{document}

\title{SnapClass: An AI-Enhanced Classroom
Management System for Block-Based Programming
\\
\thanks{This work is supported by National Science Foundation under awards 2405854 and 2405855. }
}


\author{
\IEEEauthorblockN{Bahare Riahi, Xiaoyi Tian, Ally Limke, Viktoriia Storozhevykh, Veronica Cateté, Tiffany Barnes\\
Department of Computer Science, North Carolina State University, Raleigh, NC, USA \\
\{briahi, xtian9, anlimke, vstoroz, vmcatete, tmbarnes\}@ncsu.edu\\[1ex]
Nicholas Lytle\\
College of Computing, Georgia Institute of Technology, Atlanta, GA, USA \\
nlytle3@gatech.edu\\[1ex]
Khushbu Singh\\
University of Virginia, Charlottesville, VA, USA \\
dcf2rk@virginia.edu}
}

\maketitle

\begin{abstract}
Block-Based Programming (BBP) platforms, such as
Snap!, have become increasingly prominent in K–12 computer
science education due to their ability to simplify programming
concepts and foster computational thinking from an early age.
While these platforms engage students through visual and gamified interfaces, teachers often face challenges in using them effectively and finding all the necessary features for classroom management. To address these challenges, we introduce SnapClass, a classroom management system integrated within the Snap! programming environment. SnapClass was iteratively developed drawing on established research about the pedagogical and logistical challenges teachers encounter in computing classrooms.
Specifically, SnapClass allows educators to create and customize
block-based coding assignments based on student skill levels, implement rubric-based auto-grading, and access student code history and recovery features. It also supports monitoring student engagement and idle time, and includes a help dashboard with a ``raise hand” feature to assist students in real time. This paper describes the design and key features of SnapClass those are developed and those are under progress. 

\end{abstract}


\section{Introduction}
Block Based programming (BBP) platforms have become increasingly important in computer science K-12 classes, offering students an entry point for programming concepts and developing computational thinking \cite{lee2022computer} from an early age.
Many teachers, especially those new to teaching or with limited experience  in programming, often seek support in grading and assessing student assignments \cite{milliken2019effective}. Visual programming tasks and open-ended projects add further complexity \cite{basu2019using}, as they require evaluating custom submissions that vary widely in content and structure, making consistent grading more challenging. Furthermore, teachers may apply different grading criteria to different students or groups of students even when using the same platform \cite{milliken2020gradesnap}. 
The integration of AI into learning platforms has improved accessibility and learning outcomes \cite{manoharan2021maximizing}, while reducing administrative tasks so instructors can focus more on teaching and personalized student support \cite{chen2020artificial}. Previous studies found K–12 teachers prioritized AI features such as detailed feedback, progress monitoring, personalized content support, tutoring access, and automated grading, along with fairness, plagiarism checks, customizable rubrics, and adaptable tools \cite{limke2024survey,riahi2025comparative}.

In this poster, we introduce SnapClass, a novel AI-enabled classroom management platform integrated in Snap! BPP environment. Building on prior research and teacher needs, snapClass offers AI-supported teaching and management features that address key classroom challenges. We describe the system features along with the AI-integrated feature. We also discuss its potential for future development to better serve teachers and students.
This poster is intended to be presented in the Posters and Software Demos track at VL/HCC 2026.
\section{Background}
\subsection{Learning management system }\label{AA}
Many K-12 computer science teachers have found grading programming assignments challenging due to their own lack of experience and uncertainty of how to apply rubric and high submission volume \cite{milliken2021exploring}, that lead to low confidence in grading and high pressure on them\cite{Bower2017improving}. 
Additionally, teachers’ ability to meet diverse student needs varies significantly \cite{NijenhuisVoogt2021Teaching} and simply counting coding blocks is not enough to assess the depth or sophistication of student learning, there is a critical need for targeted tools to support teachers in delivering instruction and conducting fair, consistent assessments in computer science classrooms \cite{limke2025does}. 

As identified through four case studies of secondary school teachers using an earlier version of SnapClass, displaying the student’s code, rubric, and feedback space on one screen are some of the crucial features of an effective tool. However, several challenges were unaddressed:
\begin{itemize}
  \item Automated grading- so they wouldn’t have to grade everything manually and could save more time.
  \item Real-time alerts- to quickly identify which students were having trouble and might need extra help.
  \item Personalized learning paths or adaptive curricula- so lessons and assignments can be adjusted on each student’s skill level or progress.
\end{itemize}

\subsection{Teacher- Informed Design for K-12 Auto Grading Tools}
Most auto-grading tools for block-based programming (BBP) focus on technical accuracy and are tailored to higher education, often neglecting K–12 needs like differentiated support and meaningful feedback \cite{Parihar2017,Heckman2020,Nabil2021}. SnapClass addresses this gap using evidence-based, classroom-aligned design. Through participatory methods—including co-design workshops and feedback sessions—three K–12 teachers identified key challenges and features needed for effective grading and instruction \cite{limke2023participatory,Singh2022,Khan2024}. This collaboration ensures SnapClass supports real teaching workflows and promotes fair, standards-aligned learning.
\section{SnapClass: System Design and Features}
\subsection{SnapClass: An Integrated Classroom Management System within Snap! Environment}
\begin{figure}[htp]
    \centering
    \includegraphics[width=1\linewidth]{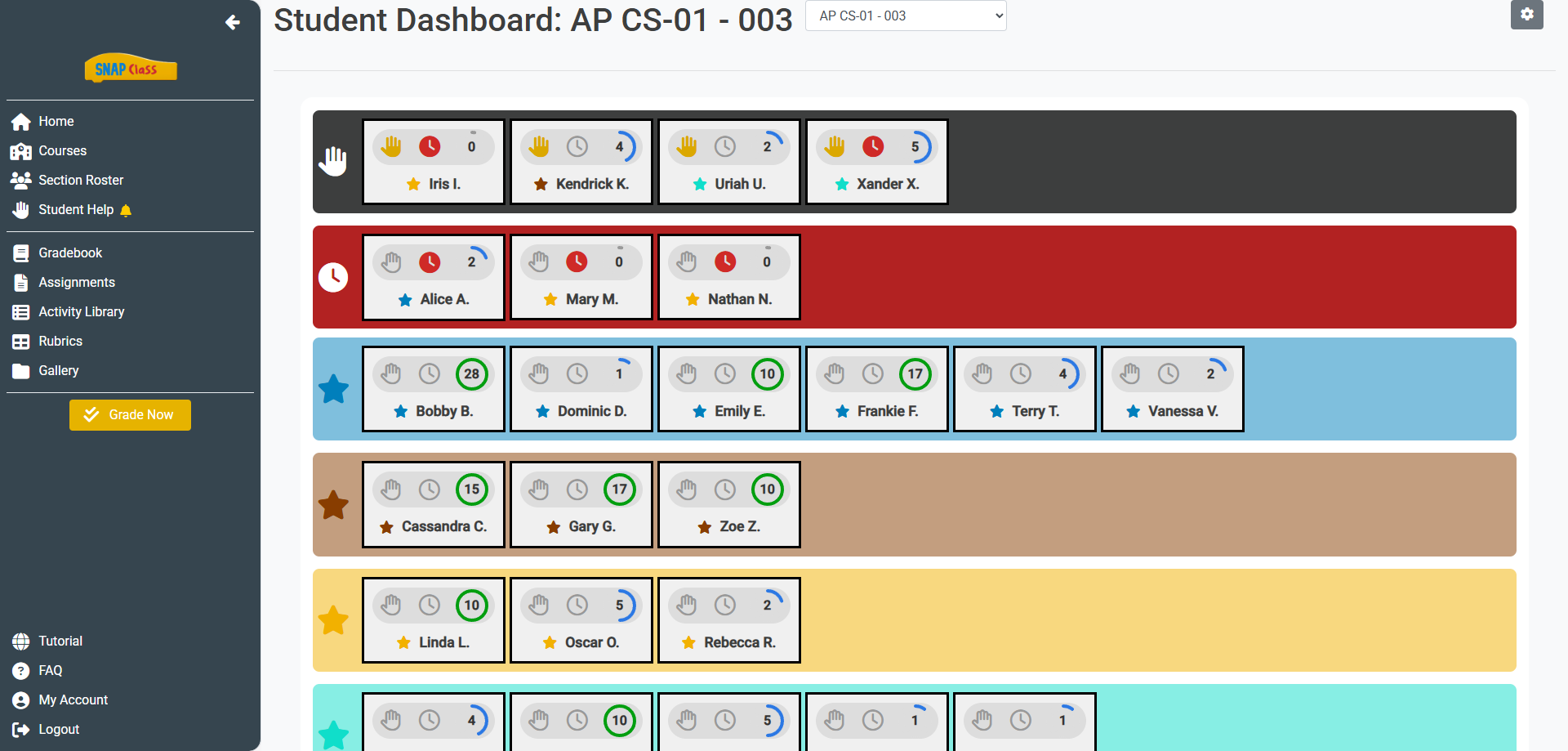}
    \caption{Teacher dashboard showing student hand raise status, activity level and coding progress}
    \label{fig:hand-raise}
\end{figure}

\begin{figure}[htp]
    \centering
    \includegraphics[width=1\linewidth]{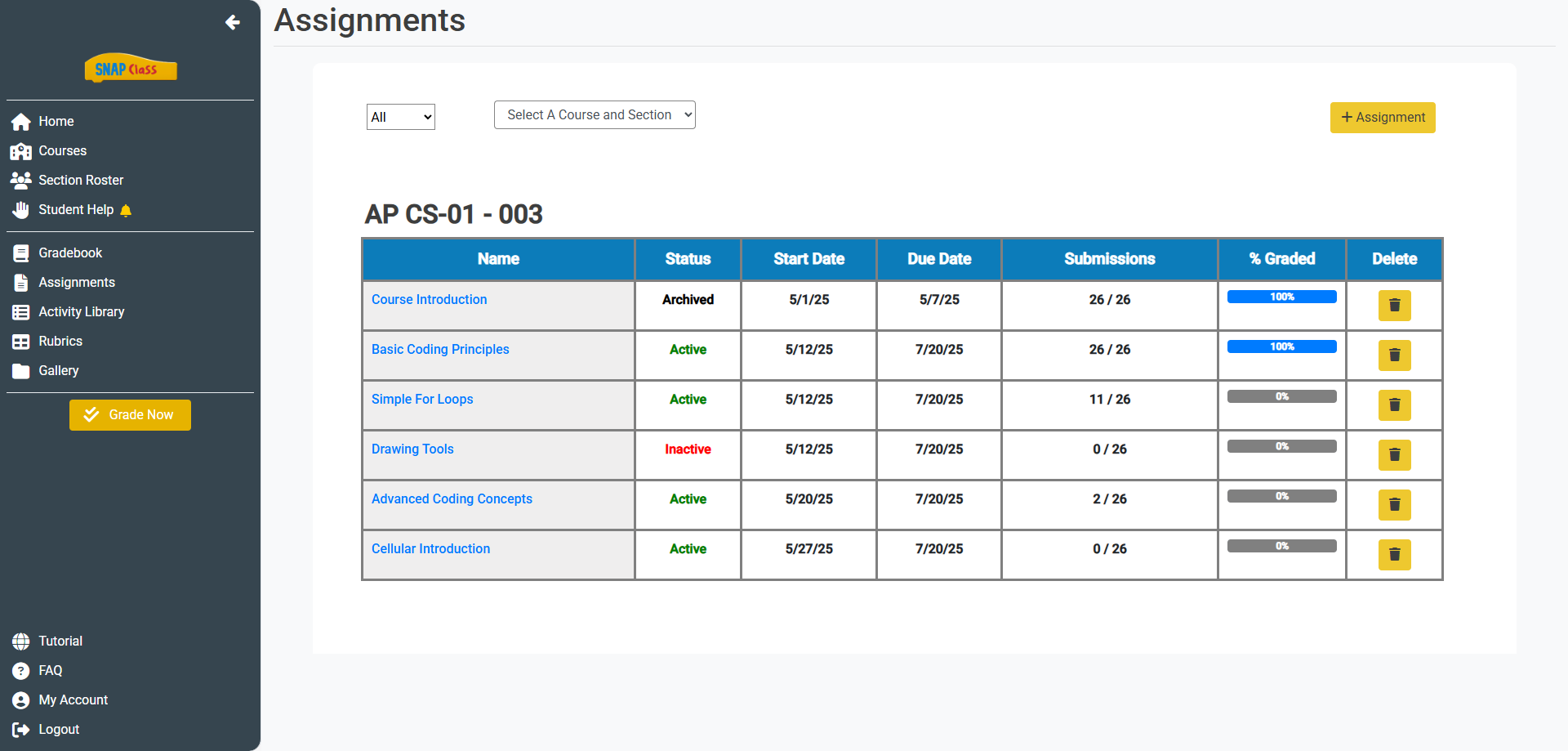}
    \caption{Teacher view of assignments}
    \label{fig:teacher-view-assignment}
\end{figure}

In this section, we present the system design and features of SnapClass. SnapClass is a classroom management system integrated within the block-based programming environment Snap!, to support teachers to seamlessly view and grade the student assignments and give feedback within the same coding platform. 
When a teacher logs into SnapClass, they will see the status of all the courses and roster of course sections, the assignment status of each class including number of submissions and grading status (Figure \ref{fig:teacher-view-assignment}). They can create new assignment and rubrics, view student coding work in the environment and record the grading. They can also monitor the summary of student activities and view students help requests in the dashboard (Figure \ref{fig:hand-raise}). 
SnapClass provides AI-based support through an integrated chatbot and auto-grading assistance. The AI chatbot is trained on the Snap manual and learning materials to offer contextual, provide relevant contextual help to students. In the grading system, a large language model (LLM)-powered rubric generator creates an initial set of rubric items for teachers to review, customize, and apply. We describe SnapClass features in the following sections. 

\subsection{Seamless Classroom Management for grading and in-class responsiveness}
\balance
 SnapClass addresses switching between platforms to view students' work and grade them\cite{riahi2025comparative}, by allowing teachers to view code, track progress, grade, and give feedback all in one interface. Its teacher dashboard also supports in-class management with features like hand raise tracking  (Figure \ref{fig:hand-raise}), idle student alerts, and progress monitoring by code blocks, enabling timely, individualized feedback.

\begin{figure}[htp]
    \centering
    \includegraphics[width=1\linewidth]{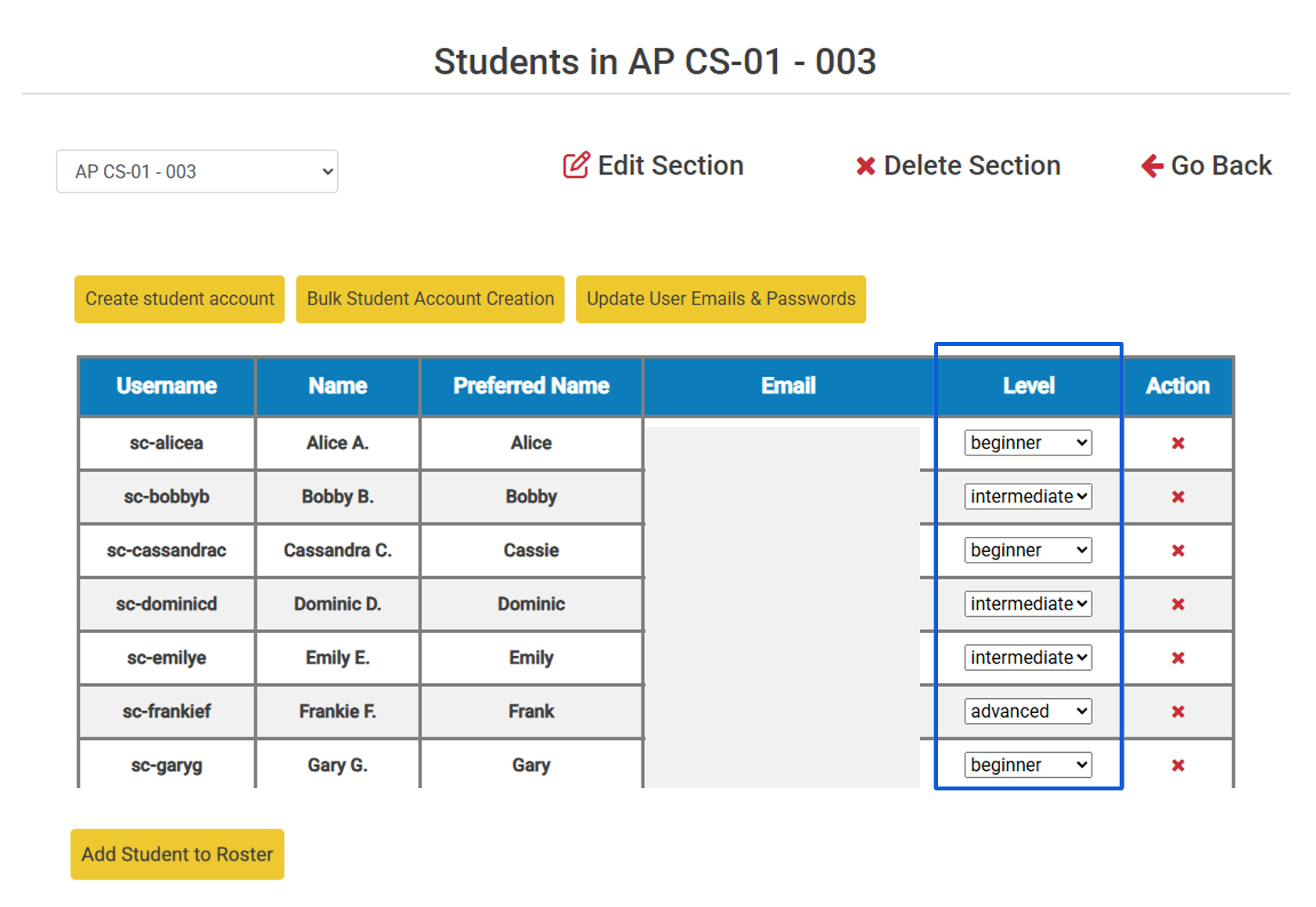}
    \caption{Student management page where teachers can set student coding levels as beginner, intermediate or advanced. Each level corresponds to different starter code in assignments.}
    \label{fig:student-management}
\end{figure}

\begin{figure}
    \centering
    \includegraphics[width=0.7\linewidth]{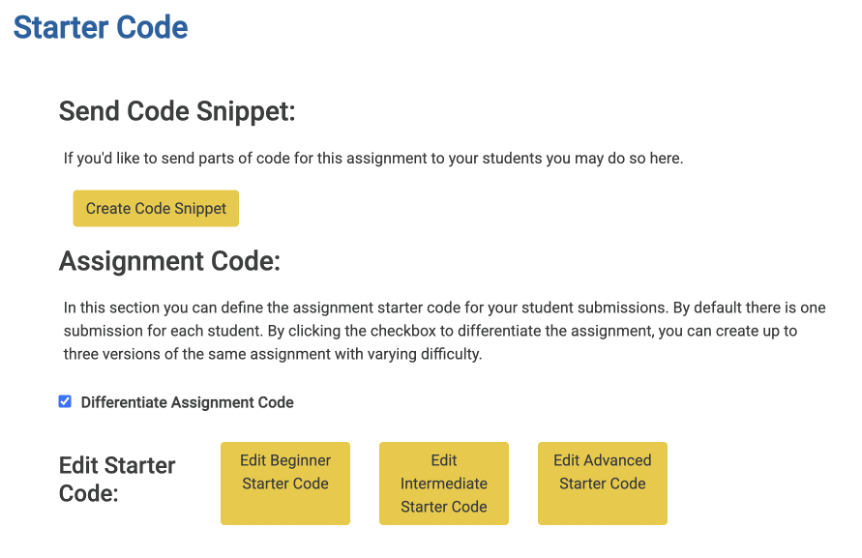}
    \caption{Assignment creation page teachers can optionally create differentiated starter code for different student levels. }
    \label{fig:assignment-creation}
\end{figure}

SnapClass supports educators to scaffold learning based on students’ prior experience and skill. In the student management page, teachers can define learners' scaffolding levels needed as beginner, intermediate and advanced (Figure \ref{fig:student-management}). Later the learners will receive different starter code for their coding assignment. Figure \ref{fig:assignment-creation} shows the assignment creation page where teachers can build differentiated starter code blocks for different levels of learners. 

\subsection{SnapBot: AI Chat for SnapClass}

\begin{figure*}[htp]
    \centering
    \includegraphics[width=\linewidth]{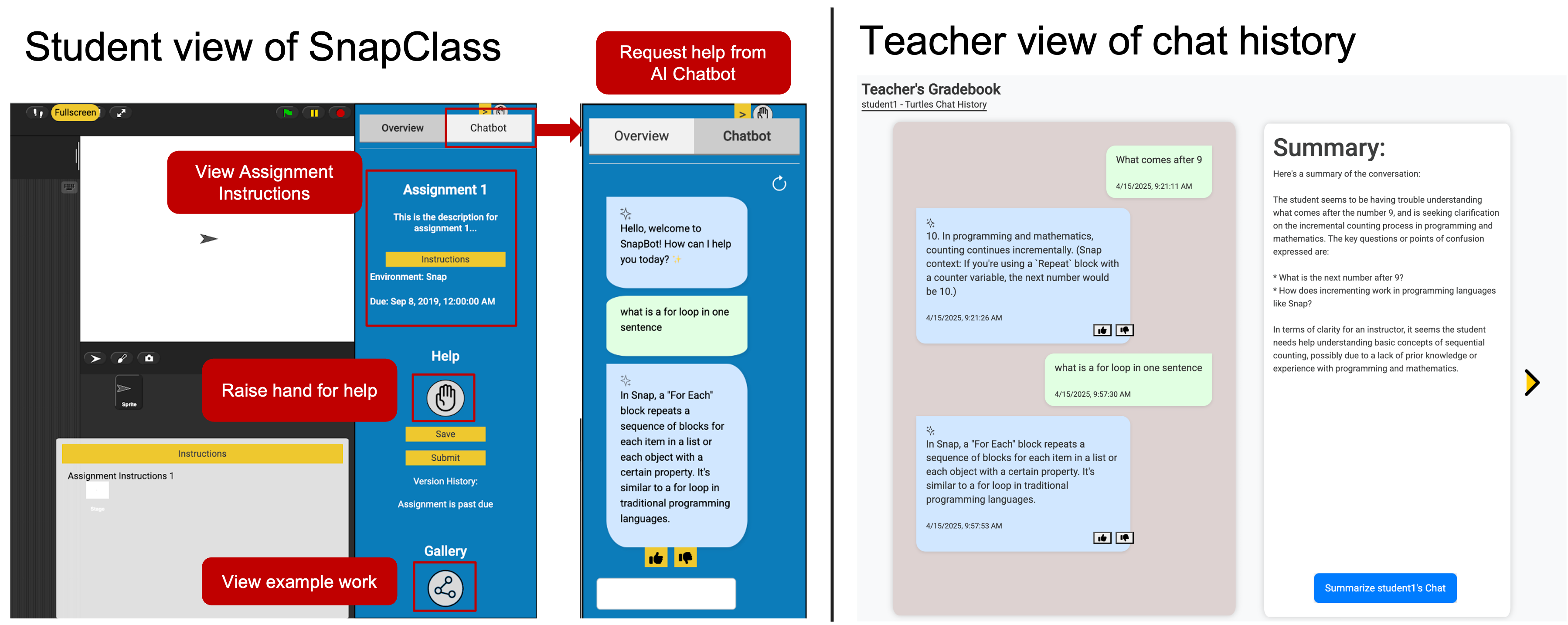}
    \caption{Left: SnapClass student view. Studnets work on Snap coding environment on the left side, and view assignment instructions, raise hand for help and chat with \textit{SnapBot}. Right: \textit{SnapBot} teacher view of individual student chat history within an assignment and a summary of the chat. }
    \label{fig:student-chatbot}
\end{figure*}

To support immediate feedback in large classrooms and reduce dependence on teacher availability, SnapClass offers students real-time, on-demand assistance through an AI chatbot called \textit{SnapBot}.\textit{SnapBot} uses a state-of-the-art LLM, Llama3.1 (70B) served on Ollama \cite{ollama2025} in combination with a retrieval-augmented generation (RAG) framework \cite{lewis2020retrieval}, which connects the LLM to additional data from the Snap Reference Manual \cite{snapManual2025}. This manual contains essential information on how to navigate the Snap environment and write block-based code, which allows the chatbot to give context-specific responses to student requests. 
Figure \ref{fig:student-chatbot} shows the student and teacher views of \textit{SnapBot}. Students access it via the “Chatbot” tab to ask questions about Snap and assignments. Teachers can enable or disable \textit{SnapBot} per assignment, hiding it when off. They can also monitor chat histories, view updated summaries every five minutes, and receive alerts for inappropriate use. Both students and teachers can rate responses, and teachers can give detailed feedback to improve chatbot accuracy and alignment with learning goals.

\subsection{AI-based auto-grading assistance}

\begin{figure}[htp]
    \centering
    \includegraphics[width=\linewidth]{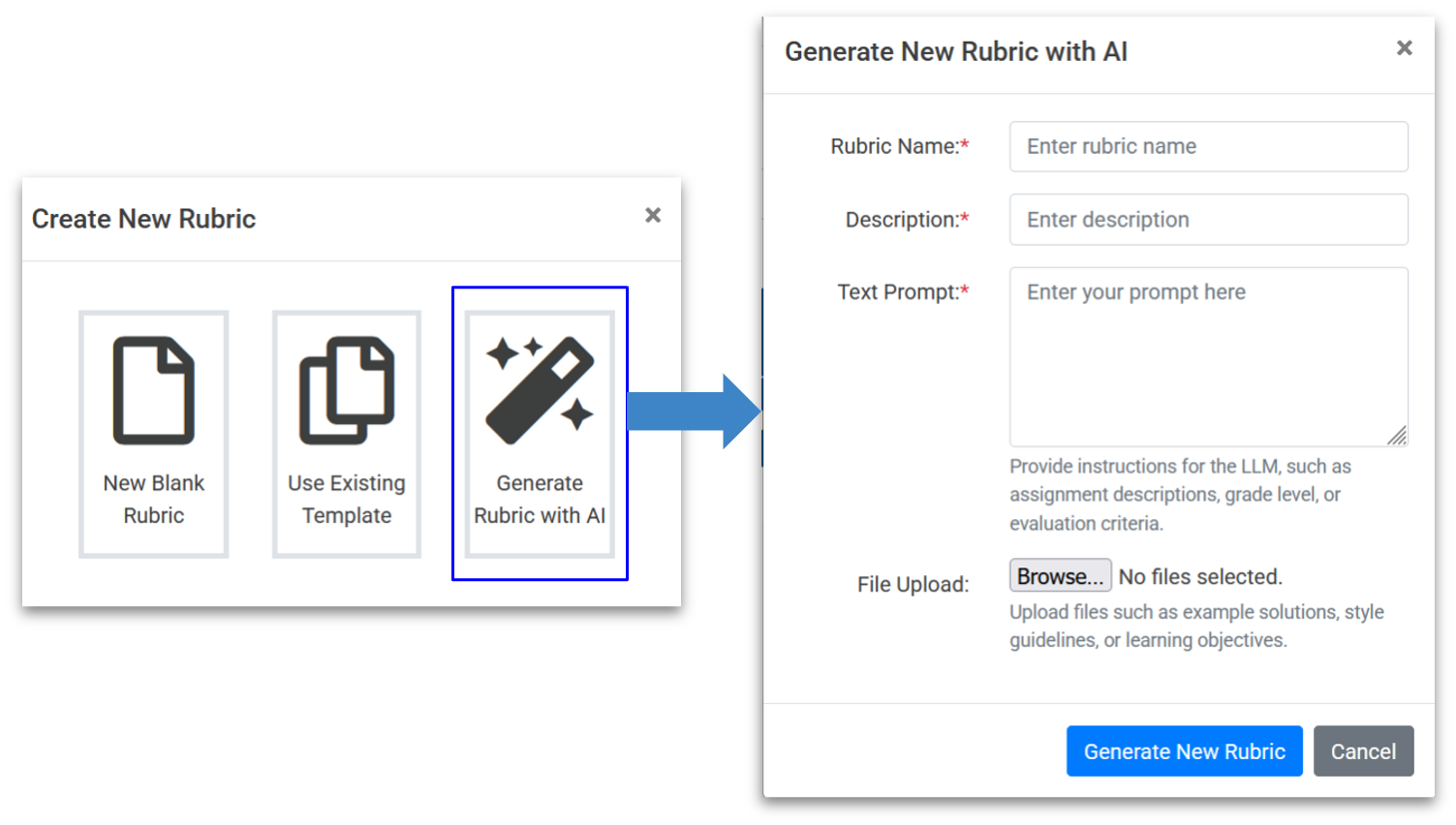}
    \caption{Rubric creation page, including options about generating rubrics with AI}
    \label{fig:rubric-creation}
\end{figure}

\begin{figure}[htp]
    \centering
    \includegraphics[width=\linewidth]{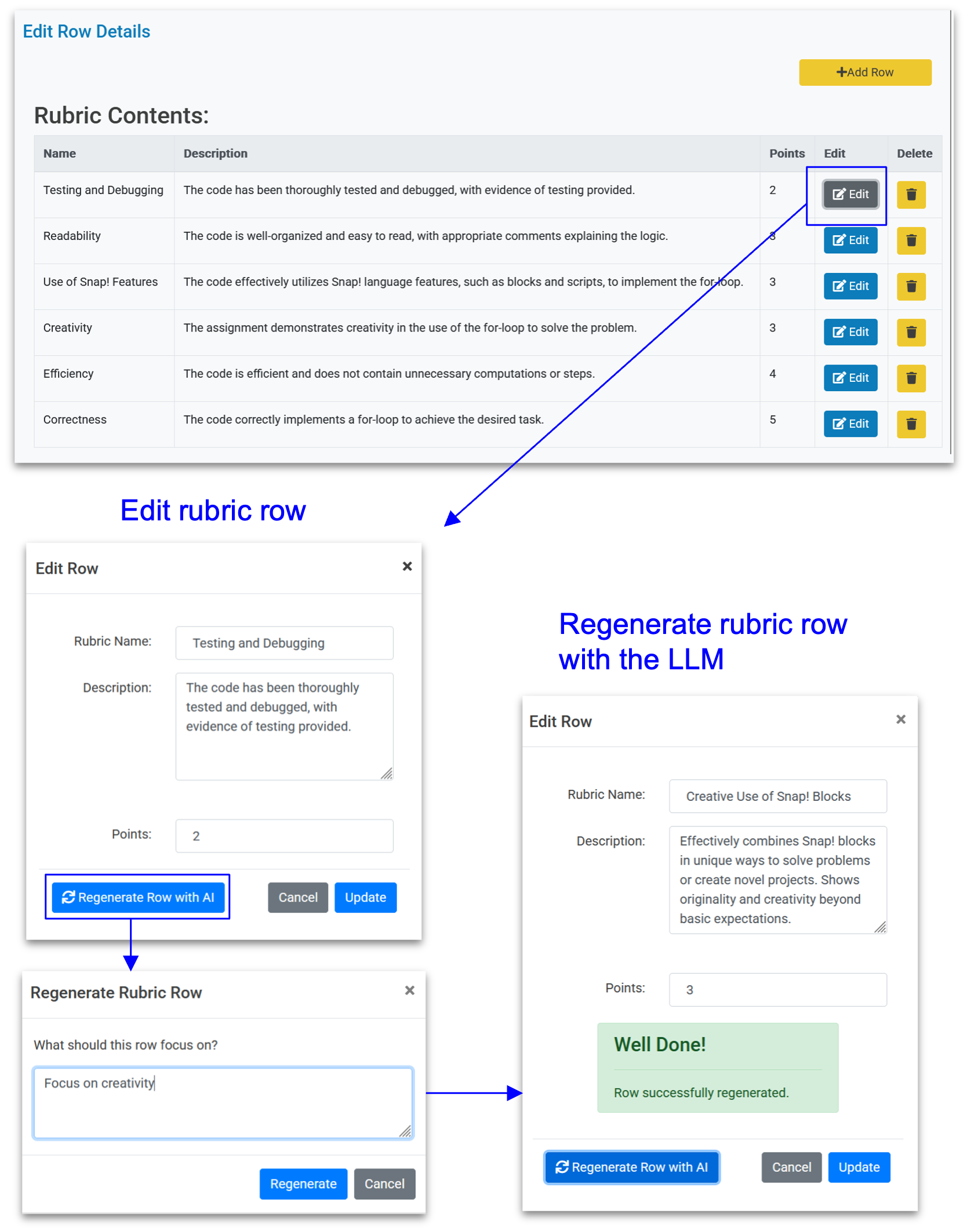}
    \caption{Rubric management page. Teachers can add, modify and delete individual rubric rows and regenerate the row with additional prompting with AI}
    \label{fig:rubric-manage}
\end{figure}

SnapClass includes an AI-based assistance for auto-grading. As shown in Figure \ref{fig:rubric-creation}, teachers can create a rubric from scratch, a template, or generate one with AI. When using AI, they provide a name, description, and prompt (e.g., grade level, assignment details, evaluation criteria), and may upload example solutions or learning objectives. The system then generates a rubric (Figure \ref{fig:rubric-manage}) that teachers can edit, refine, or save for future use.

\section{Discussion and Future Work}
 SnapClass was designed to bridge the gap between the practical needs of K–12 educators and the capabilities of existing BBP environments. The aim of integrating snapclass features with AI is to provide scalable, real-time support for both instruction and assessment, while reducing teacher workload and maintaining pedagogical flexibility. The SnapBot chatbot empowers students to access just-in-time support, especially helpful in large classrooms or asynchronous learning environments. 
Future work will include empirical studies with more teachers and students, expanding AI capabilities to support formative feedback, and refining chatbot personalization for varied learning needs.

\newpage


\vspace{12pt}
\color{red}

\end{document}